\documentclass[prl,superscriptaddress,showpacs,amssymb,amsmath,
amsfonts,aps,twocolumn,secnumarabic,floatfix]{revtex4}
\usepackage{graphics}
\usepackage{epsfig}
\begin{document} 
\bibliographystyle{try} 
 
\topmargin 0.1cm 
 
 \title{Three-body mechanisms in the $^3$He(e,e$'$p) reactions at high missing momentum}

\newcommand*{\SACLAY }{ CEA-Saclay, Service de Physique Nucl\'eaire, F91191 Gif-sur-Yvette, Cedex, France} 
\affiliation{\SACLAY } 

\newcommand*{\JLAB }{ Thomas Jefferson National Accelerator Facility, Newport News, Virginia 23606} 
\affiliation{\JLAB } 

\author{J.M.~Laget}
     \affiliation{\SACLAY}
     \affiliation{\JLAB}

\date{\today} 
 
\begin{abstract} 

A particular three-body mechanism is responsible for the missing strength which has been reported in $^3$He(e,e$'$p) reactions at missing momentum above 700 MeV/c. It corresponds to the absorption of the virtual photon by a nucleon at rest which subsequently propagates on-shell and emits a meson which is reabsorbed by the pair formed by the two other nucleons. Its amplitude, which is negligible in photon induced reactions as well as in the electro-production of an on-shell meson, becomes maximal in the quasi-free kinematics ($X=1$). It relates the amplitude of the $^3$He(e,e$'$p)D reaction to the amplitude of $pD$ elastic scattering at backward angles. 
  
\end{abstract} 
 
 
\maketitle

\begin{figure}[hbt]
\begin{center}
\epsfig{file=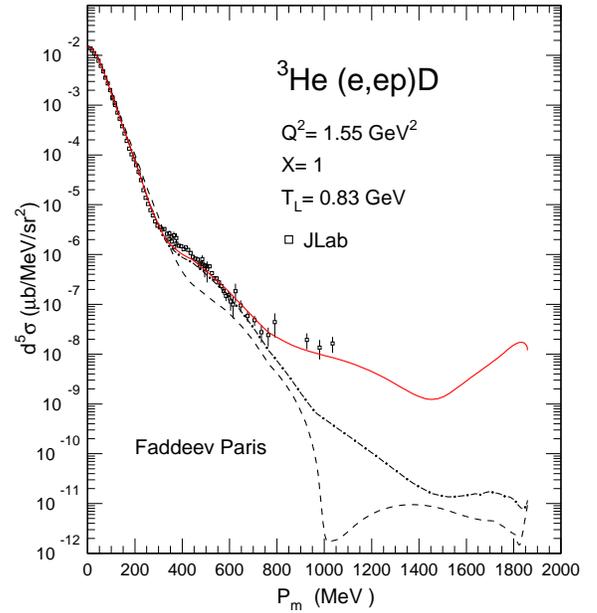,width=3.0in}
\caption[]{(Color on line). The momentum distribution in the $^3$He(e,e$'$p)d reaction at $E_e=4.795$~MeV, X=1 and
$Q^2=1.55$~GeV$^2$. Dashed line: Plane Wave.  Dash-dotted line: 2 body FSI, MEC and $\Delta$ production included. Full line: 3 body mechanisms included. The data are from ref.\cite{Mar04}.}
\label{he3pd}
\end{center}
\end{figure}

The recent study at JLab of the  $^3$He(e,e$'$p) reactions~\cite{Mar04,Fat04} represents a text book result. One-body, two-body and three-body mechanisms have been disentangled in a single experiment. The two-body break-up channel, Fig.~\ref{he3pd}, is particularly illuminating in this respect. One-body mechanisms, where the electron interacts with a single nucleon and the deuteron is spectator (Plane Wave), dominate below missing momentum $p_m\simeq$ 300~MeV/c (the momentum of the outgoing deuteron). For missing momenta between 300 and 700~MeV/c, nucleon-nucleon Final State Interactions (FSI) and Meson Exchange Currents (MEC) between two nucleons, take over~\cite{La04}. However, the recombination of one of these two active nucleons with the third nucleon, to form the outgoing deuteron, strongly reduces the corresponding cross section at high $p_m$: Independent evaluations~\cite{La04,Sch04} of the contributions of one and two-body mechanisms miss by one order of magnitude the experiment at $p_m\simeq$ 1~GeV/c. Here, three-body mechanisms dominate and restore the agreement with data. The first reason is that sequential meson exchanges is the more economical way to share the large momentum transfer among the three nucleons. The second reason is that the on-shell nucleon singularity maximizes the contribution of one of these diagrams in the quasi-elastic kinematics ($X=Q^2/2m\nu=1$, $Q^2$ and $\nu$ being respectively the virtuality and the energy of the virtual photon, while $m$ is the mass of the nucleon), for the same reason~\cite{La04} why the contribution of nucleon-nucleon rescattering is maximized. The virtual photon is absorbed  by a nucleon at rest which subsequently propagates on-shell before emitting a meson which is reabsorbed by the two other nucleons. In real photon induced reactions~\cite{La88,Dh89}, the corresponding singularity is suppressed by the high momentum components of the nuclear wave function since, for kinematical reasons, the absorption by a nucleon at rest is forbidden. Also, the intermediate nucleon can not be on-shell in the electro-production of an on-shell meson. The corresponding three nucleon amplitude has been switched off~\cite{La88} in my previous evaluation~\cite{La04}. This note is devoted to the proper evaluation of the contribution of this diagram and its on-shell singularity.

The method~\cite{La81} is based on the expansion of the reaction amplitude in terms of few relevant diagrams, which are computed in the momentum space, in the Lab. frame. The kinematics as well as the propagators are relativistic and no angular approximation is made in the evaluation of the loop integrals. The elementary operators which appear at each vertex have been calibrated against the corresponding channels. Its application  to the $^3$He(e,e$'$p) channels  has been discussed in refs.~\cite{La85,La87,La88}, and updated in refs.~\cite{La04} to the JLab energy range. A comprehensive summary is given in ref.~\cite{La94} for the $^4$He(e,e$'$p)T channel. 

For the sake of the discussion, I reproduce the expression of the amplitude of the three-body mechanisms~\cite{La88} for the $^3$He(e,e$'$p)D reaction and the main steps of its evaluation: 
\begin{eqnarray}
T= i\int \frac{d^3\vec{n}}{(2\pi)^3} \sum_{m_n} T(\gamma ^3He\rightarrow npp)
\nonumber \\
\times  \left [(\frac{1}{2}m_p\frac{1}{2}m_n | 1m_2) 
\frac{U^D_0(\vec{n}-\vec{p_2}/2)}{\sqrt{4\pi}}
\;\; +\;\; \mathrm{D}\; \mathrm{ Wave} \right]
 \label{pd}
\end{eqnarray}
where $U^D_0$ and $U^D_2$ are the $S$ and $D$ components of the final deuteron wave function for the Paris potential~\cite{PaXX}, and where $(\vec{n},m_n)$ and $(\vec{p},m_p)$ are respectively the momenta and magnetic quantum numbers of the neutron and the proton which form the final deuteron. The four momenta of the incoming photon and the outgoing proton and deuteron are respectively $(\nu,\vec{k})$, $(E_1,\vec{p_1})$ and $(E_2,\vec{p_2})$. Only the positive energy part of the nucleon propagators is retained, and the integral over the energy of the neutron picks the pole in its propagator and puts it on shell.

The three-body break-up amplitude $T(\gamma ^3He\rightarrow npp)$ corresponds to a two-loop diagram and the integral runs over the the four momenta of the two nucleons which do not absorb the photon. Again the energy integration picks  their poles and put them on-shell. After changing the variables of integration we are left with a six fold integration over the relative $(\vec{q})$ and the total $(-\vec{p'})$ momenta of these two nucleons:
\begin{eqnarray}
T(\gamma ^3He\rightarrow npp)= -\sqrt{3} \sum_{\Lambda} 
(1\Lambda \frac{1}{2}m'_p|\frac{1}{2} m_i) \nonumber \\
\int \frac{d^3 \vec{p'}}{(2\pi)^3} \frac{1}{\sqrt{4\pi}}
\frac{\chi_0^{T=0}(\vec{p'})}{q^2_{\pi}-m^2_{\pi}+i\epsilon}  
T_{\gamma p}(\vec{k},\vec{p'}m'_p \rightarrow \vec{q_{\pi}},\vec{n}m_n)
\nonumber \\
T_{\pi^+(np)_0}(\vec{q_{\pi}},-\vec{p'}\Lambda \rightarrow \vec{p}m_p,\vec{p_1}m_1)
\label{ppn}
\end{eqnarray}
where $(\vec{p'},m'_p)$ and $(-\vec{p'},\Lambda)$ are respectively the momenta and the magnetic quantum numbers of respectively of the proton on which the pion is photo-produced and the pair which absorbs it.

\begin{figure}[hbt]
\begin{center}
\epsfig{file=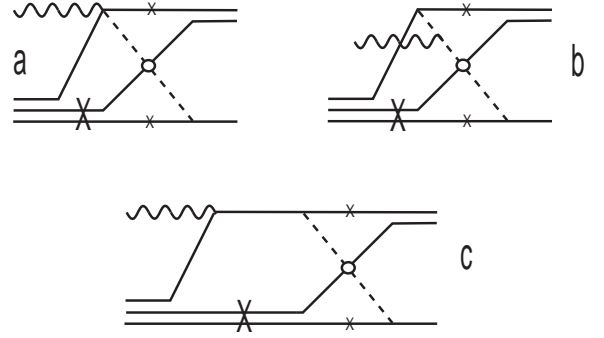,width=3.0in}
\caption[]{The three-body graphs. The nucleons which are put on-shell by the integration over the energies are marked by a cross. Dashed lines represent $\pi$.}
\label{graphs}
\end{center}
\end{figure}

The full antisymmetrized three-body wave function~\cite{Haj81} is the solution of the Faddeev equation for the Paris potential. It is expanded on a basis where two nucleons couple to angular momentum $L$, spin $S$ and isospin $T$, third nucleon moving with an angular momentum $l$. A very good approximation~\cite{La87} is to factorize each component into a product, $\Phi^T_{Ll}= U^T_L(q)\chi^T_l(p)$, of two wave functions which describe respectively the relative motion of the two nucleons inside the pair and the motion of the third. Since pion absorption by a $T=1$ pair is strongly suppressed~\cite{An86,Ba85}, only absorption by a $T=0$ pair is retained. This prevents the formation of the $\Delta$ in the pion electro-production elementary amplitude, since the total isospin of the $pD$ final state is $T=1/2$. The three relevant Born terms lead to the three-body graphs which are depicted in Fig.~\ref{graphs}.

In the contact (a) and the photoelectric (b) graphs, which were retained in ref.~\cite{La04,La88}, the only singularity comes from the pole in the pion propagator. I refer to~\cite{La88} for the details of its treatment and of the choice of the matrix elements of the various elementary amplitudes which enter Eq.~\ref{ppn}. The nucleon Born term (c) exhibits an additional singularity which needs a different treatment. After expanding the pion electro-production amplitude $T^B_{\gamma p\rightarrow n\pi^+}$ and retaining the nucleon Born term, the corresponding part of the three-body break-up amplitude takes the form:
\begin{eqnarray}
T^N(\gamma ^3He\rightarrow npp)=-\frac{3}{2} \frac{e g_{\pi NN}\sqrt{6}}{2m} \sum_{\Lambda} 
(1\Lambda \frac{1}{2}m'_p|\frac{1}{2} m_i) \nonumber \\
G_M^p(Q^2)\left [\frac{(m_n |\vec{\sigma}\cdot\vec{q_{\pi}} \vec{\sigma}\cdot\vec{k}\times\vec{\epsilon} |m'_p)}{q^2_{\pi}-m^2_{\pi}}
\right. \nonumber \\ \left.
 T_{\pi^+(np)_0}(\vec{q_{\pi}},0 \Lambda \rightarrow \vec{p}m_p,\vec{p_1}m_1) \right ]_{\vec{p'}=0}
\nonumber \\
\int \frac{d^3 \vec{p'}}{(2\pi)^3} \frac{1}{\sqrt{4\pi}}
\frac{\chi_0^{T=0}(\vec{p'})}{2E_N(p^0 _N- E_N+i\epsilon)}\;\;  
\label{ppn_N}
\end{eqnarray}
where the pion propagator, the pion absorption amplitudes and the spin-momentum part of the of nucleon Born amplitude~\cite{BL77,La88a} have been factorized out the integral which can be expressed in a compact analytic form according to ref~\cite{La81}. Besides $\pi^+$ exchange, $\pi^{\circ}$ exchange contributes also and the combination of the two amplitudes leads to the factor $3/2$. The on-shell and off-shell energies of the intermediate nucleon are respectively $E_N= \sqrt{(\vec{p'}+\vec{k})^2 + m^2}$ and $p^0_N= \nu + m_{^3He} - \sqrt{\vec{p'}^2 +4m^2}$. Eq.~\ref{ppn_N} displays the Transverse (magnetic) part of the current. Its Longitudinal (Coulomb) part exhibits a similar form but contributes weakly to the differential cross section at the high virtuality $Q^2$ relevant to this experiment.  

\begin{figure}[hbt]
\begin{center}
\epsfig{file=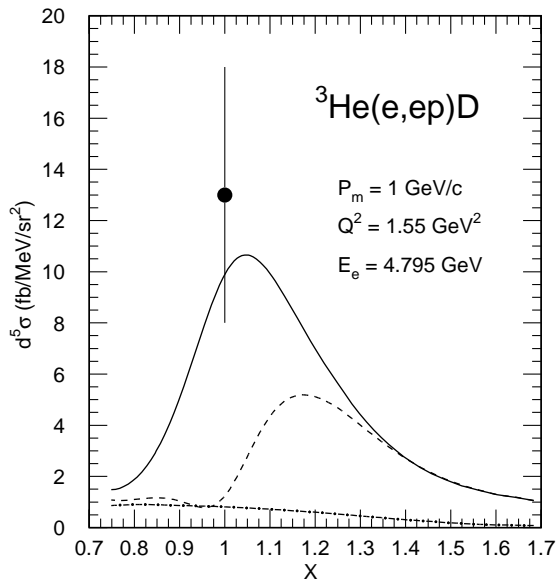,width=3.0in}
\caption[]{The angular distribution of the deuteron emitted with a constant momentum $p_m= 1$~GeV/c in the $^3$He(e,e$'$p)d reaction when $Q^2=1.55$~GeV$^2$ and $E_e=4.795$~GeV. Dash-dotted line: without nucleon Born term (c) in the three-body amplitude.  Dashed line: principal part of the nucleon Born term included. Full line: singular and principal part of the nucleon Born term included. The datum is from ref.\cite{Mar04}.}
\label{xdis}
\end{center}
\end{figure}

As in the NN scattering amplitude~\cite{La04}, the singular (on-shell) part of the integral exhibits a sharp maximum, while its principal (off-shell) part strictly vanishes, in the quasi-elastic kinematics at $X=1$ (Fig.~\ref{xdis}). Here, the virtual photon is absorbed by a proton at rest in $^3$He which propagates on-shell before emitting a (off-shell) pion which is reabsorbed by the pair, formed by the two other nucleons, also at rest in $^3$He. This justifies the factorization of the elementary matrix elements, as well as the pion propagator, in eq.~\ref{ppn_N}  and their evaluation assuming that the active nucleon and the spectator deuteron are at rest in $^3$He. This is an accurate and parameter free way to evaluate  the nine-fold integral (eq.~\ref{pd}) since it relies on the low-momentum component of the $^3$He wave function and on-shell elementary matrix element which have been calibrated independently against the corresponding reactions. For instance, eqs.~\ref{pd}, \ref{ppn} and~\ref{ppn_N} relate the two-body disintegration of $^3$He to backward pD elastic scattering. From studies performed at Saturne and Dubna, it is dominated by the exchange of an interacting $\pi$N pair (see {\it e.g.} ref~\cite{La93}). Away from $X=1$, the nine-fold integral must be evaluated completely, but the three-body amplitude is not dominant and contributes at the same level as others: one enters into the land of uncertainties (off-shell extrapolations, high-momentum components, etc\ldots) which I avoid in this note.

Not only the magnitude of the three-body, as well as of the two-body, amplitudes are well under control in the quasi elastic kinematics, but also the sharing of the momentum transfer between the three nucleon is responsible for the flattening of the cross section at high missing momenta in Fig.~\ref{he3pd}. The rise at the highest missing momentum reflects the increasing probability of forming a fast deuteron with two fast nucleons: this is only possible because their momentum mismatch is strongly reduced in the three-body mechanisms, contrary to one- and two-body mechanisms (where one nucleon at least is almost spectator). The experimental confirmation of the high missing momentum part of the distribution is highly desirable. This can be best achieved by detecting the fast deuteron in coincidence with the electron (the Lorentz boost focuses more deuterons than protons in the detector). A preliminary results~\cite{Vou04} confirms the prediction at $p_m\simeq$ 1.20~GeV/c, and confirms the trend of the (e,e$'$p) data. A new experiment may use a third large acceptance magnetic spectrometer~\cite{BB} in Hall A at JLab. One may also dig into the data already available at CLAS~\cite{Me03} in Hall B at JLab, if the statistics permits. Also, the determination of the angular distribution (Fig.~\ref{xdis}) of the fast deuteron at constant missing momentum $p_m\sim$ 1~GeV/c will map out the characteristic rapid variation of the three-boby amplitude against $X$.

The virtual photon may also couple directly to two active nucleons which propagate before reinteracting with the third. However, such a mechanism is suppressed by the electromagnetic form-factor of the pair, which falls down faster than the nucleon form factor when the virtuality $Q^2$ increases. Also, at $X=1$, the kinematics prevents the proton to be emitted at rest and consequently the photon interacts with a moving pair. A measure of these two effects is the strong reduction of the small backward peak (highest $p_m$) with respect to the forward peak ($p_m=0$) in the Plane Wave curve of  Fig.~\ref{he3pd}. On the contrary, the corresponding contributions are maximized at $X=2$ since the proton can be emitted at rest. Their experimental study remains to be done at high $Q^2$.  

\begin{figure}[hbt]
\begin{center}
\epsfig{file=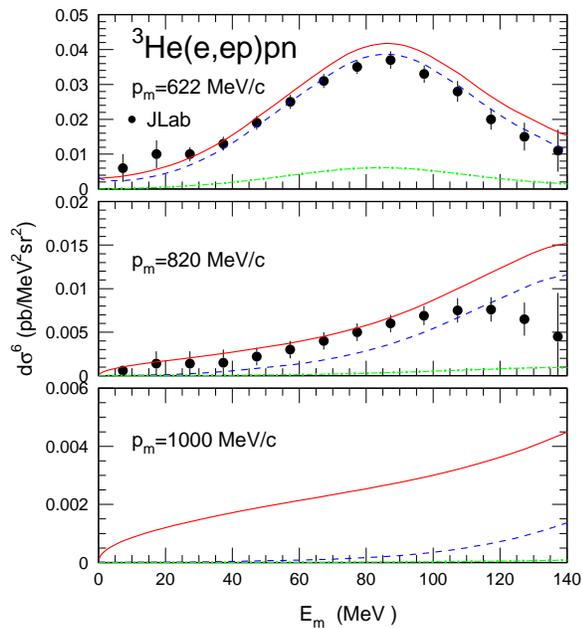,width=3.0in}
\caption[]{(Color on line). The missing energy distributions in the $^3$He(e,e$'$p)np reaction at X=1 and $Q^2=1.55$~GeV$^2$ for selected values of large missing momenta. Dash-dotted lines: PW.  Dashed lines: 2 body FSI, MEC and $\Delta$ production included. Full lines: 3 body mechanisms included. The data are from ref.\cite{Fat04}.}
\label{cont}
\end{center}
\end{figure}

Turning to the continuum, Fig.~\ref{cont} shows typical missing energy (the internal energy $E_m$ of the undetected $pn$ pair) spectra at high recoil momentum (the total momentum $p_m$ of the $pn$ pair) in the $^3$He(e,e$'$p)pn reaction~\cite{Fat04}. At $p_m=600$~MeV/c and below, the peak in the continuum emphasizes two nucleon mechanisms~\cite{La04,Cio04}: Its top corresponds to the kinematics of the disintegration of a pair of nucleons at rest, while its width reflects the Fermi motion of the spectator nucleon. Again, on-shell nucleon-nucleon rescattering dominates in the quasi-free kinematics (X=1) achieved in this experiment. At higher missing momentum, this two-nucleon peak appears on top of a broad continuum, which is dominated by the nucleon Born part (Fig.~\ref{graphs}c) of three-body mechanisms. It has been implemented according to eq.~\ref{ppn_N} in the continuum three-body amplitude~\cite{La88b}: besides the two other Born terms (a and b), it contains also the $\Delta$ formation term which is not forbidden by isospin selection rules. When integrated over the missing energy up to the pion threshold, the contribution of the three-body mechanisms reconciles  the theoretical momentum distribution with the experimental one and gets rid of the discrepancy in Fig.~3 of ref.~\cite{Fat04} at high missing momenta.  

In Fig.~\ref{cont}, the missing energy spectra have been plotted up to the pion production threshold, since the experiment has been performed with two magnetic spectrometers. To go beyond, one needs to detect two nucleons in coincidence with the electron. This can be achieved with a third spectrometer~\cite{BB} in Hall A or with CLAS~\cite{Me03} in Hall B at JLab.

The extension to the $^4$He disintegration channels~\cite{An04} would also be interesting since the relative importance of the two- and three-body mechanisms is expected to be different from $^3$He. For instance, the extra form factor in the amplitude of pion photo-production on a pair of nucleons (see Fig.~4 and eq.~5 in ref.~\cite{La94}) suppresses the contribution of the nucleon pole in the three-body mechanisms. But the presence of a fourth nucleon allows other mechanisms of which the amplitudes remain to be investigated.

In summary, this note completes the study of the electrodisentegration of $^3$He which I undertook several years ago. As soon as the recoil momentum increases, Nature prefers to share the momentum transfer first between two nucleons and then between the three nucleons, rather to transfer it to a single nucleon. This picture is on solid quantitative grounds in the quasi-elastic kinematics ($X=1$) where the singularity associated with the propagation of an on-shell nucleon maximizes the corresponding amplitudes: they depend on low momentum components of the wave function and on on-shell elementary operators. It reproduces, without any free parameter over six decades, the beautiful text-book experiment which has recently been performed at JLab. Far from the quasi-elastic kinematics, one enters into the domain where the tails of these amplitudes overlap and become less under control.  One may wonder whether high momentum components of the wave function will ever be accessible.

\end{document}